# Calibrating transition metal energy levels and oxygen bands in first principles calculations: accurate prediction of redox potentials and charge transfer in lithium transition metal oxides

Dong-Hwa Seo (서동화), Alexander Urban, and Gerbrand Ceder*

Department of Materials Science and Engineering, Massachusetts Institute of Technology, Cambridge, Massachusetts 02139, United States.

*Corresponding author*: gceder@mit.edu

**ABSTRACT**

Transition metal oxides play an increasingly important role in technology today including applications such as catalysis, solar energy harvesting, and energy storage. In many of these applications, the details of their electronic structure near the Fermi level are critically important for their properties. We propose a first-principles based computational methodology for the accurate prediction of oxygen charge transfer in transition metal (TM) oxides and lithium TM (Li-TM) oxides. To obtain accurate electronic structures, the Heyd-Scuseria-Ernzerhof (HSE06) hybrid functional is adopted and the amount of exact Hartree-Fock exchange (mixing parameter) is adjusted to reproduce reference band gaps. We show that the HSE06 functional with optimal mixing parameter yields not only improved electronic densities of states but also better energetics (Li-intercalation voltages) for $LiCoO_2$ and $LiNiO_2$ as compared to GGA, GGA+$U$ and standard HSE06. We find that the optimal mixing parameters for TM oxides are system-specific and correlate with the covalency (ionicity) of the TM species. Strong covalent (ionic) nature of TM-O bonding leads to lower (higher) optimal mixing parameters. We find that optimized HSE06 functionals predict stronger hybridization of the Co $3d$ and O $2p$ orbitals than GGA, resulting in a greater contribution from oxygen states to charge compensation upon delithiation in $LiCoO_2$. We also find that the band gaps of Li-TM oxides increase linearly with the mixing parameter, enabling the straightforward determination of optimal mixing parameters based on GGA ($\alpha = 0.0$) and HSE06 ($\alpha = 0.25$) calculations. Our results also show that $G_0W_0$@GGA+$U$ band gaps of TM oxides (MO, M = Mn, Co, Ni) and $LiCoO_2$ agree well with experimental references, suggesting that $G_0W_0$ calculations can be used as a reference for the calibration of the mixing parameter in case no experimental band gap has been reported.

**INTRODUCTION**

Charge transfer (CT) between a transition metal (TM) atom and its ligands sensitively affect the properties of materials for various applications related to energy storage,[1-5] electrocatalysts,[6,7] optical materials,[8] magnetic materials,[9,10] and superconducting materials.[11] Thus, many efforts have been made to quantify and predict selective CT between TM atoms and coordinating species computationally and experimentally. Zaanen *et al.* first introduced CT to classify TM oxides as CT insulators and Mott-Hubbard (MH) insulators.[12] The authors find that if the CT energy ($\Delta$) from filled oxygen $p$ orbitals to empty TM $d$ orbitals is smaller than the Coulomb and exchange energy ($U_{dd}$) between TM $d$ orbitals in the TM oxides, electronic excitations are mainly determined by CT.[12,13] The band gaps of such CT insulators are proportional to $\Delta$. In contrast, if $\Delta$ is larger than $U_{dd}$, TM oxides act as MH insulators and their band gaps are proportional to $U_{dd}$. Compounds in which $\Delta$ is similar to $U_{dd}$ are mixed type of CT and MH insulators, which are in the intermediate region of the Zaanen-Sawatzky-Allen (ZSA) phase diagram.[12] van Elp *et. al.* experimentally observed with Photoelectron spectroscopy (PES) and Bremsstrahlung Isochromate spectroscopy (BIS) that the magnitude of $\Delta$ and $U_{dd}$ is similar in late TM monoxides such as MnO and CoO.[14,15] Both, valence and conduction bands of these TM oxides have strongly mixed TM $3d$ and O $2p$ character, which confirms the intermediate nature between CT and MH insulators.

The issue of what the lowest excitation in TM oxides is has recently taken on particular relevance in energy storage applications, as evidence of preferential ligand oxidation over transition metal oxidation has emerged, creating potentially a novel mechanism by which charge can be stored in Li-ion batteries. Oxygen redox activity has been proposed as a possible source of the excess lithium extraction capacity in Li-excess-TM-oxide intercalation materials, such as $Li_2MnO_3$-$LiMO_2$,[16,17] $Li_2Ru_{1-y}M_yO_3$ (M = Mn, Sn, Ti)[4,18,19], Co doped $Li_2O$[20] $Li_xNi_{2-4x/3}Sb_{x/3}O_2$[21] and Li-Nb-M-O[22,23] systems. Such Li-excess-TM-oxides are technologically appealing as cathode materials in lithium ion batteries. In conventional Li-TM intercalation cathodes, such as $LiMn_2O_4$[24] and $LiFePO_4$,[25] the TM is oxidized upon lithium extraction and reduced upon lithium reinsertion. However, the aforementioned materials exhibit a surplus capacity that cannot be explained by the TM redox couples but is commonly attributed to oxygen redox activity.[4,16-20,22] Reversible charge transfer to oxygen in bulk electrode materials may become an exciting new pathway for energy storage with increased energy density. Therefore, a reliable methodology to investigate CT between oxygen, lithium and TM atoms is a requirement to understand which TM oxides facilitate reversible oxygen oxidation.

In fact, the contribution of oxygen to the redox activity of conventional Li intercalation materials has already been addressed by computations and experiments. The local CT from lithium atoms to their neighboring oxygen atoms in $Li_{1-x}CoO_2$ with lithiation (lithium insertion) has been demonstrated by first principles calculation in the mid 90's.[1,2,26] Experimental results from X-ray absorption spectroscopy (XAS) and Electron energy loss spectroscopy (EELS) of $LiCoO_2$ confirmed these predictions.[27,28] Whitingham *et al.* proposed a mechanism whereby both, TM and oxygen, are involved in the charge compensation during charge/discharge in $LiNi_{1/3}Mn_{1/3}Co_{1/3}O_2$.[5] Hence the redox potentials of these Li-TM-oxides are directly related to the CT on the oxygen as well as TM ions with lithium de/intercalation.

The examples in the previous paragraphs underline the importance of understanding CT phenomena in TM oxides. Unfortunately, a quantitative computational investigation of CT requires a very accurate prediction of the electronic structure of the TM oxide. Density Functional Theory (DFT)[29] on the level of the generalized gradient approximation (GGA) cannot predict the electronic structure of TM oxides with the required accuracy, as the self-interaction error (SIE) results in an over-delocalization of the electron.[30-34] In semiconductors PBE is known to misalign the metal $3d$ states and the chalcogenide $p$ states.[35-38]. Furthermore, GGA significantly underestimates the band gaps of TM monoxides.[39,40] Introducing a Hubbard-$U$ correction (GGA+$U$)[30-32] for the TM $d$ orbitals artificially localizes the electrons on the TM atoms, but not on the oxygen atoms,[30-32,41] and GGA+$U$ still underestimates the band gaps of TM monoxides.[39,40] Admixing exact Hartree-Fock (HF) exchange makes it possible to correct the SIE in both, TM and oxygen, simultaneously.[42-44] Therefore, hybrid functionals that explicitly add a fraction of the HF exchange energy, $E_x^{\mathrm{HF}}$, to the GGA exchange-correlation energy,

$$E_{xc}^{\mathrm{GGA}} = E_x^{\mathrm{GGA}} + E_c^{\mathrm{GGA}}, E_{xc}^{\mathrm{hybrid}} = (1-\alpha)E_x^{\mathrm{GGA}} + \alpha E_x^{\mathrm{HF}} + E_c^{\mathrm{GGA}}, \quad (1)$$

are the natural choice to study band and state alignment. Indeed, band gaps of TM monoxides calculated with the HSE06 hybrid functional[45-47] are larger than those calculated with GGA and GGA+$U$.[39,40] However, the band gaps of MnO and CoO, to name just two examples, still differ from the experimental ones by more than 0.7 eV.[39,40,48] If an error of similar magnitude is to be expected for the relative position of the TM $d$ and oxygen $p$ valence bands, no quantitative conclusions regarding the amount of CT in these systems would be possible.

The amount of exact HF exchange admixed to a GGA functional is determined by the mixing parameter α of equation (1), and adjusting this parameter provides a further handle to improve the accuracy of the electronic structure. Two main approaches to determine the mixing parameter have been described in the literature:

(1) **Empirical fitting**: the mixing parameters for the B3PW91 and B3LYP functionals were chosen to reproduce thermochemical properties in Pople's Gaussian database (G1).[42,49]

(2) **Perturbation theory**: the mixing parameter used in the Perdew–Burke–Ernzerhof hybrid functional (PBE0) and the Heyd−Scuseria−Ernzerhof (HSE) functional (an approximation of the former) is 25%, which was determined analytically via perturbation theory.[45,50,51] Subsequent benchmarks showed that HSE predicts accurate thermochemical properties for molecular test sets (G2),[45] and good band gaps for simple semiconductors such as C, Si, BN, BP, SiC, $\beta$-GaN, GaP, and MgO with a mean absolute error (MAE) of 0.2 eV, which is much better than either LDA and PBE  (MAE: ~1.4 eV).[46]

In this article, we show that empirical adjustment of the HSE mixing parameter to reproduce optical band gaps and the electronic density of states (DOS) obtained from highly accurate electronic structure calculations can significantly improve the description of CT effects in TM oxides and Li-TM-oxides. As such, HSE with optimal mixing parameter becomes a predictive approach for the accurate description of electrochemical and electronic structure properties of TM oxides, thereby making it an important tool for the study and design of the next generation of energy devices.

The HSE functional with the default mixing parameter ($\alpha = 0.25$) significantly overestimates the redox potentials of Li-TM-oxides (LiCoO$_2$ and LiNiO$_2$), as it too strongly localizes the electrons on oxygen atoms.[43] By adjusting the mixing parameter to reproduce experimental band gaps (determined from PES-BIS experiments), this artificial electron localization can be reduced. Since few PES-BIS results for TM oxides have been reported, we propose to predict band gaps of Li-TM-oxides using highly accurate GW approximation (GWA) calculations,[52] which can then be used as reference to determine suitable mixing parameters.

Such empirical adjustment of the mixing parameter by itself is not a new idea. Han *et al*. reported adjusting HSE mixing parameters for oxides to experimental band gaps.[53] Graciani *et al*. obtained optimal mixing parameters for $CeO_2$ and $Ce_2O_3$ through a fitting to experimental band gaps.[54] Siegel *et al*. recently determined mixing parameters for $Li_2O_2$ by fitting the GWA band gap.[55] However, unlike the previous studies which focused on particular applications, we seek to establish a general and robust methodology for the derivation of optimal system-specific mixing parameters, and an assessment of the accuracy of this approach.

In the following section, the computational set-up and the mixing parameter adjustment is discussed in detail. Subsequently, optimal mixing parameters for TM oxides and Li-TM-oxides are determined. Finally, HSE calculations using optimized mixing parameters are applied to investigate the redox potentials of various Li-TM-oxides.

**METHODS**

Spin-polarized generalized gradient approximation (GGA) calculations were carried out based on the PBE exchange-correlation functional.[56] Projector-augmented wave (PAW) pseudopotentials were used as implemented in the Vienna *ab initio* simulation package (VASP).[57] We employed a plane-wave basis set with a kinetic energy cutoff of 520 eV for the representation of the wavefunctions and a *gamma* centered $8 \times 8 \times 8$ k-point grid for the Brillouin zone integration. The atomic positions and lattice parameters of all structures were optimized until residual forces were smaller than 0.02 eV/Å. Rhombohedral $2 \times 2 \times 2$ supercells containing eight formula units of MO were used for MO (M = Mn, Co, Ni) and were fully relaxed with antiferromagnetic spin ordering and local ferromagnetic spin ordering in the [111] direction, as observed in experiments.[39] In the case of lithium cobalt oxide, the hexagonal primitive cell of O3-Li$_{1-x}$CoO$_2$ with R3-m space group was used.[58] A monoclinic primitive cell with C2/m space group was used for LiNiO$_2$ to allow for the Jahn-Teller distortion of the Ni-O bond.[59] The rotationally invariant scheme by Dudarev *et al.*[60] was used for the Hubbard $U$ correction to GGA ($GGA+U$). For the TM oxides, the $U$ values from reference[61] were employed ($U$[Mn$^{2+}$] = 3.9 eV, $U$[Co$^{2+}$] = 3.4 eV, and $U$[Ni$^{2+}$] = 6.0 eV), which were fitted to the experimental binary formation enthalpies. For M$^{3+}$ and M$^{4+}$ in Li$_x$MO$_2$, the self-consistently calculated $U$ values for TM ions in layered structures were used ($U$[Co$^{3+}$] = 4.9eV, $U$[Co$^{4+}$] = 5.4eV, $U$[Ni$^{3+}$] = 6.7eV, $U$[Ni$^{4+}$] = 6.0eV).[41] The average voltages of LiMO$_2$ were calculated with average $U$ values of M$^{3+}$ and M$^{4+}$.

The HSE screened Coulomb hybrid density functional introduces exact HF exchange to the PBE exchange-correlation functional. The HSE exchange-correlation energy is defined as

$$E_{xc}^{\text{HSE}} = \alpha E_x^{\text{HF, short}}(\mu) + (1 - \alpha) E_x^{\text{PBE, short}}(\mu) + E_x^{\text{PBE, long}}(\mu) + E_c^{\text{PBE}}, \quad (2)$$

where $E_x^{HF}$ and $E_x^{PBE}$ are the exact HF and PBE exchange energies, respectively, $E_c^{PBE}$ is the PBE correlation energy, $\alpha$ is the mixing parameter and $\mu$ is a range-separation parameter.[45,62] The HSE functional is split into short-range (short) and long-range (long) terms to exclude the slowly decaying long-range part of the HF exchange. HSE06 employs a range-separation parameter of $\mu = 0.207 \text{ Å}^{-1}$, which results in a reasonable compromise between accuracy and computational cost.[63] For each TM oxide, we sampled mixing parameters within the range $0 \leq \alpha \leq 0.5$ to fit the reference band gaps.

When experimental band gaps were not available, many-body perturbation theory in the GWA was employed to accurately estimate band gaps.[52] In the GW approximation, Hedin's equations[64] for the quasi particle (QP) energy are solved by a first-order expansion of the self-energy operator in the one-body Green's function (G) and the screened Coulomb interaction (W). The non-self-consistent $G_0W_0$ approximation was previously reported to predict accurate band gaps for TM oxides.[65,66] Our $G_0W_0$ calculations were based on initial wavefunctions and eigenvalues obtained from GGA+$U$ calculations, thus we denoted these calculations as $G_0W_0$@GGA+$U$. The usual random-phase approximation (RPA) was employed to calculate the dielectric matrix for the screened Coulomb interaction.[64] For all GW calculations, we used a plane-wave energy cutoff of 347 eV and 128 bands (i.e., more than 100 unoccupied bands), which was confirmed to converged band gaps for TM oxides and Li-TM-oxides.

To further confirm the accuracy of the bonding interactions between TM and oxygen atoms in TM oxides, computational oxygen K-edge EELS spectra were compared to experimental references. The $Z$+1 approximation was employed to calculate EELS spectra with GGA+$U$ and HSE06,[67,68] which required large supercells of $4 \times 4 \times 4$ primitive unit cells for TM oxides and $3 \times 3 \times 3$ primitive unit cells for $LiCoO_2$. For these supercells, a *gamma*-centered $1 \times 1 \times 1$ k-point grid was used.

**RESULTS**

**1. Optimizing the HSE mixing parameter for TM oxides**

As discussed in the previous section, we optimized the mixing parameters ($\alpha$) of TM monoxides (MnO, NiO and CoO) and layered Li-TM-oxides (Li$_{1-x}$CoO$_2$ and Li$_{1-x}$NiO$_2$, $x$ = 0 and 1) by fitting reference band gaps from PES-BIS[14,15,69] and GWA calculations. In principle, the *band gap* is the difference in energies between the highest occupied valence band and lowest unoccupied conduction band. However, a direct comparison of computed band gaps with experimentally measured values is difficult, due to the intrinsic instrumental resolution and the resulting broadening of spectra. Therefore, we adjusted the mixing parameter to match the shape of the calculated DOS to PES-BIS spectra after reducing the resolution of the computed DOS intensities by convolution with Gaussian distributions. Since valence and conduction bands are observed by different spectroscopical techniques (PES and BIS, respectively) that exhibit different instrumental broadenings and intensities,[70] Gaussian distributions with different full width at half maximum (FWHM) were adopted: an FWHM of 1 eV was used for valence band states, and 2 eV for conduction band states, respectively. The intensities of the valence and conduction bands of the calculated DOS were also rescaled individually in order to match the PES and BIS spectra. In all DOS calculations, the valence band maximum was shifted to zero. Both, PES and BIS spectra, were simultaneously shifted to align the offset of the PES spectra to the DOS valence band.

When using G$_0$W$_0$@GGA+$U$ as reference, we directly compared the actual band gaps, i.e., the energy difference between valence and conduction band, without any additional broadening.

## 1) TM oxides (MO, M = Mn, Co and Ni)

**Figure 1** shows a comparison of the PES-BIS spectra for MnO, NiO, and CoO from reference 14, 15, 62 with the DOS as calculated with GGA, GGA+$U$, HSE06, and $G_0W_0$ (all DOS and PES-BIS spectra are aligned as described above). For each of the three oxides, the onset of the GGA and GGA+$U$ conduction bands occurs at several eV lower energy than observed by BIS, indicating that GGA and GGA+$U$, in agreement with previous reports,[39,40] significantly underestimate the band gaps of TM oxides. HSE06 with standard mixing parameter ($\alpha = 0.25$) slightly underestimates the band gap of MnO by 0.7 eV, overestimates the one of CoO by 0.65 eV, but accurately predicts the band gap of NiO. Excellent agreement between HSE06 and experimental reference is achieved by adjusting the mixing parameters for MnO and CoO to 0.30 and 0.20, respectively. We also compared the band gap calculated with $G_0W_0$@GGA+$U$ to the experimental reference. As can be seen in **Figure 1**, $G_0W_0$@GGA+$U$ calculations predict the experimental band gaps and peak shapes of MnO, NiO and CoO well with an accuracy of about ± 0.5 eV, which corresponds to an uncertainty of approximately ± 0.04 in the mixing parameter. Hence, $G_0W_0$@GGA+$U$ can be used as reference method when experimental data is not available.

Our band gaps calculated with HSE06 and $G_0W_0$@GGA+$U$ (without broadening) are in good agreement with previous computational reports.[39,40,48] Note that if the actual computational band gap (i.e., the difference between valence and conduction band edges) is compared to the experimental "band gap", it would appear as if HSE06 and $G_0W_0$@GGA+$U$ dramatically underestimate the band gap of MnO (as previously reported[39]). However, this is an artifact caused by a small gap state around 2~3 eV above the Fermi level, which is not visible in the BIS spectrum (**Figure 1a**). Even though the HSE06 and $G_0W_0$@GGA+$U$ band gaps of MnO are apparently smaller than the experimental one, the shape of the DOS matches well with the experimental spectrum. These results point out that our method of matching the peak onsets of the broadened computed DOS with the experimental spectra is more robust than the direct comparison of the band gaps, and they additionally confirm again that $G_0W_0$@GGA+$U$ predicts band gaps of TM oxides well.

To further evaluate the accuracy of the oxygen $2p$ states in MnO and NiO and their hybridization with the TM states, O K-edge EELS spectra were calculated (**Figure 2**). The first peak (A) in the spectra originates from the hybridized oxygen $2p$ and TM $3d$ bands and the second and third peaks (B and C) are related to the hybridized oxygen $2p$ and TM $4s/p$ states.[71,72] Calculated and measured spectra were aligned at the first peak (A). The relative peak positions and peak ratio of O K-edge EELS spectra of MnO and NiO calculated with HSE06 are in better agreement with the experimental reference than those from GGA+$U$ calculations, especially near the Fermi energy. Note that HSE06 successfully predicts the peak between A and B in the O K-edge EELS spectra of MnO, which was not assigned in a previous experimental report,[71] whereas GGA+U fails to predict this peak, as shown in **Figure 2**.

### 2) Li-TM-oxides (LiMO$_2$, M = Co and Ni)

The same procedure was applied to Li-TM-oxides. As expected from the TM monoxides evaluated in the previous section, GGA underestimates the band gap of LiCoO$_2$ (**Figure 3a**). However, the GGA+$U$ DOS reproduces the features of the PES-BIS spectrum with reasonable accuracy and the band gap is only about 0.5 eV lower than in experiment (**Figure 3a**). HSE06 with standard mixing parameter ($\alpha = 0.25$) significantly overestimates the band gap of $LiCoO_2$ (4.0 eV vs. 2.7 eV),[15] and a much lower mixing parameter ($\alpha = 0.17$) is required to obtain the correct result (**Figure 3a**). Note that the optimal mixing parameter for $LiCoO_2$ ($\alpha = 0.17$) is lower than the one found for CoO ($\alpha = 0.20$). Also shown in **Figure 3a** is the $G_0W_0$@GGA+$U$ (U = 4.9 eV) DOS, which well predicts the peaks of the experimental spectra and the band gap of $LiCoO_2$. The difference between the $G_0W_0$@GGA+$U$ and the experimental band gap is less than 0.3 eV, which translates to a variation of $\pm 0.02$ in the mixing parameter of $LiCoO_2$.

The O K-edge EELS spectra of $LiCoO_2$ calculated with GGA+$U$ and HSE with the optimal mixing parameter are nearly identical and are in good agreement with the experimental reference (**Figure 3b**).[27] The first sharp peak at ~2eV in the EELS spectrum of **Figure 3b** is related to the hybridized state of oxygen $2p$ and Co $3d$ orbitals, and the broad feature between 8 and 15 eV originates from the hybridization of oxygen $2p$ and Co $4sp$ orbitals.[27,28,73] Both states are well predicted by HSE06 with the optimal mixing parameter.

The redox potential of an intercalation cathode is a function of the relative energy of the material's lithiated and (partially) delithiated phases. Therefore, an accurate description of both end points is necessary. To reveal differences in the mixing parameters of the oxides with different degree of oxidation, and to quantify the dependence of the band gap on the fraction of exact HF exchange, we calculated the band gaps of $LiCoO_2$ and $LiNiO_2$, as well as their delithiated phases $CoO_2$ and $NiO_2$, using the HSE functional with mixing parameters within the range $0 \leq \alpha \leq 0.5$. For this exercise, we consider as *band gap* the exact energy difference between the valence band and conduction band edges. For all four materials, the band gap increases linearly with the mixing parameter, as shown in **Figure 4a** and **4b**. The band gap of delithiated $CoO_2$ is much smaller than that of $LiCoO_2$ at the same mixing parameter (**Figure 4a**). The linear dependence of the band gap on the fraction of HF exchange enables the rapid determination of optimal mixing parameters by extrapolation based on the PBE ($\alpha = 0.0$) and HSE06 ($\alpha = 0.25$) data points.

Since no experimental PES-BIS reference for $CoO_2$, $LiNiO_2$, and $NiO_2$ is available, the mixing parameters for these systems were adjusted to fit the $G_0W_0$@GGA+$U$ band gaps. The optimal mixing parameter of delithiated $CoO_2$ is 0.24, which is significantly larger than that of $LiCoO_2$ ($\alpha = 0.17$). The optimal mixing parameters for $LiNiO_2$ and $NiO_2$ are 0.18 and 0.25, respectively.

## 2.   Predicted voltages of $LiCoO_2$ and $LiNiO_2$ with optimal mixing parameters

Besides electronic properties we also evaluate energy quantities. The Li-extraction voltage from Li-TM-oxides, is exactly defined as the change in energy with Li concentration,[2,3] can be measured with very high accuracy and depends sensitively on the energy of the level from which the compensating electron is extracted. As such it is an ideal quantity to calibrate electronic structure methods. The voltage of $LiMO_2$ (M = Co, Ni) can be obtained from DFT energies[2,3] as

$$V = -\frac{E(Li_{x_1}MO_2) - E(Li_{x_2}MO_2) - (x_2 - x_1)E(Li)}{(x_2 - x_1)F}, (3)$$

where $E(Li_xMO_2)$ and $E(Li)$ are the DFT energies of $Li_xMO_2$ and bcc Li metal, respectively, and $F$ is the Faraday constant. As previously reported, the average voltage of $Li_{1-x}CoO_2$ within $0 \leq x \leq 1$ is 3.38 V for GGA, 3.85 V for GGA+$U$, and 4.51 V for standard HSE06,[43] as compared to the experimental voltage of 4.1 V.[74] Thus, GGA and GGA+$U$ underestimate the average voltage, whereas HSE06 overestimates it. Using the optimal mixing parameter of the previous section, i.e., $\alpha = 0.17$ for LiCoO$_2$, the average HSE voltage becomes 4.19 V, which is in good agreement with the experimental reference (**Figure 5**). However, using instead the optimal mixing parameter of delithiated CoO$_2$ ($\alpha = 0.24$, almost equal to the standard mixing parameter) yields a much higher average voltage of 4.42 V (**Figure 5**). As can be seen in **Figure 5**, the average voltage increases linearly with the fraction of exact HF exchange for mixing parameters within $0 \leq \alpha \leq 0.3$.

In order to predict the voltage profile of Li$_x$CoO$_2$, the energies of the intermediate phases of Li$_{1-x}$CoO$_2$ (for $x = 0.75$, 0.66, 0.50, 0.33, 0.25, R3-m space group) were calculated with GGA, GGA+$U$ and HSE06. The stable Li/vacancy orderings of these intermediate phases have previously been reported for the GGA functional.[75,76] **Figure 6** shows the Li$_{1-x}$CoO$_2$ voltage profiles calculated with GGA, GGA+$U$ and HSE06 (for $\alpha = 0.17$ and $\alpha = 0.25$). The voltage profile calculated with HSE and using the optimal mixing parameter of LiCoO$_2$ ($\alpha = 0.17$) agrees well with the experimental one.[77] Although the standard HSE06 ($\alpha = 0.25$) functional results in an overall similar voltage profile, it significantly overestimates the voltage of Li$_{1-x}$CoO$_2$, in particular for $0.33 \leq x \leq 0.66$. Note that despite underestimating the average voltage by about 1 V, GGA predicts similar steps to HSE06 results in the voltage profile. GGA+$U$, on the other hand, predicts a wrong voltage profile without any steps and much lower average voltage than the experimental reference, as none of the intermediate phases are predicted to be stable by GGA+$U$ ($U = 5.1$ eV). It is noteworthy that the voltage of Li$_{1-x}$CoO$_2$ at $0.66 \leq x \leq 1$ predicted with the standard HSE06 is better agreement with experimental one than that predicted with the optimal mixing parameter of LiCoO$_2$ ($\alpha = 0.17$). The observations that HSE with low mixing parameter and uncorrected GGA predict the behavior of $Li_xCoO_2$ qualitatively well are consistent with the fact that it is among the few metallic-like Li-TM-oxides[78] when sufficient carriers are created.[79] Though the strong rise of voltage of $Li_{1-x}CoO_2$ for $x \rightarrow 1$ is more consistent with a localized hole character on oxygen, also reflected in a contraction of the O-O distance.[26] Reproducing the proper electronic structure and energetics at this high state of oxidation therefore requires a higher degree of exact exchange.

The analysis of the average voltage for the corresponding nickel compound, $Li_{1-x}NiO_2$ ($0 \leq x \leq 1$), is shown in **Figure 7**. Again we find that the optimal mixing parameter of the lithiated material (i.e., $\alpha = 0.18$ for $LiNiO_2$) yields an average voltage that is close to the experimentally observed one (3.85 V vs. 3.9 V),[80] whereas the optimal mixing parameter of delithiated $NiO_2$ ($\alpha = 0.25$) yields a much higher average voltage. As in the case of the cobalt compound, the average voltage of $Li_{1-x}NiO_2$ within $0 \leq x \leq 1$ linearly increases with the amount of exact HF exchange energy within $0 \leq \alpha \leq 0.3$ (**Figure 7**).

### 3. Band alignment in $LiCoO_2$

To better understand the impact of exact HF exchange on the electronic structure in general and on CT in particular, we compare the projected density of states (pDOS) of the Co $3d$ orbitals and O $2p$ orbitals in $LiCoO_2$ with various mixing parameter. As can be seen in **Figure 8**, the pDOS of the Co $3d$ orbitals (black lines) and the O $2p$ orbitals (red lines) in the energy range from 0 eV to -2 eV have similar shapes even though the intensity of the Co $3d$ pDOS is in general larger than the intensity of the O $2p$ pDOS. This is because the valence states are composed of hybridized states between the Co $3d$ orbitals and O $2p$ orbitals ($t_{2g}$ states). **Figure 9** shows the integrated ratio of O $2p$ orbitals to Co $3d$ orbitals component in the energy range from 0 eV to -2 eV as a function of the mixing parameter. The ratio of O $2p$ orbitals to Co $3d$ orbitals in that energy range increases with the mixing parameter, indicating that hybridization between O $2p$ orbital to Co $3d$ orbitals becomes stronger with greater mixing parameter. Note that GGA+$U$ predicts a greater O $2p$/Co $3d$ ratio than HSE with the standard mixing parameter ($\alpha = 0.25$), because the Hubbard $U$ term of GGA+$U$ stabilizes (i.e., lowers the energy of) the Co $3d$ states, which results in a stronger overlap between the Co $3d$ and O $2p$ states. All pDOS plots in **Figure 8** have been aligned at the valence band maximum (Fermi level, E = 0 eV). The alignment to the Fermi level has the same effect on the HSE results.

### DISCUSSIONS

The electronic structure predicted by DFT/GGA and DFT/GGA+U is not accurate enough to draw conclusions about the charge transfer between oxygen and TM atoms. Standard GGA is well known to overly delocalize electrons and, as its self-interaction correction depends on the orbital delocalization, it cannot properly describe the energy difference between very different orbitals such as the $3d$ TM and oxygen $2p$ states. While GGA+U removes self-interaction in the $3d$ TM orbitals, thereby allowing them to localize, it does not correct the oxygen states. As a result, GGA and GGA+U do not properly describe the electronic structure and energy of those Li-TM-oxides ($Li_xMO_2$) that exhibit strong CT, yielding unreliable redox potentials. Admixing exact Hartree-Fock (HF) exchange, i.e., using hybrid functionals, generally improves the electron localization on oxygen and TM atoms or their hybridized orbitals. The degree of localization is determined by the amount of exact HF exchange defined by the mixing parameter in hybrid functionals: the larger the fraction of exact HF exchange is, the more localized is the charge. We demonstrated that the optimal amount of HF exchange can be determined by adjusting the hybrid-functional mixing parameter to reproduce reference band gaps, i.e., experimental or GW band gaps. **Figure 2 and 3** show that the HSE hybrid functional with optimized mixing parameter reproduces experimental O K-edge EELS spectra of TM oxides and Li-TM-oxides very well, and for TM oxides, O K-edge EELS spectra calculated with HSE are in much better agreement with experimental results than GGA+$U$ results, especially near the Fermi energy. The fact that these mixing parameters optimized to reproduce electronic structure also significantly improve the energetics of oxidation, as described by the Li-extraction voltage, is encouraging and supports the idea that the optimized HSE functionals describe the bonding in these materials better.

The optimal mixing parameter is system specific (**Table 1**) reflecting differences in the nature of the TM-oxygen interaction. For both Li-TM-oxides, LiCoO$_2$ ($\alpha = 0.17$) and LiNiO$_2$ ($\alpha = 0.18$), the optimal mixing parameters are lower than those of the corresponding TM monoxides ($\alpha = 0.20$ for CoO and $\alpha = 0.25$ for NiO). It is known from PES-BIS spectroscopy that the covalency of LiCoO$_2$ is stronger than that of CoO,[15,81] and stronger covalency induces less charge localization on the TM and oxygen atoms, thus demanding a lower fraction of exact exchange. Note that the degree of the covalency of the TM-oxygen bond is inversely proportional to the charge transfer energy $\Delta$.[15] Previously reported values for $\Delta$ are 4.0 eV for LiCoO$_2$,[15] 5.5 eV for CoO,[15] 6.2 eV for NiO,[82] and 8.8 eV for MnO,[14] which exhibit exactly the same trend as the optimized mixing parameters (0.17 for LiCoO$_2$, 0.20 for CoO, 0.25 for NiO, and 0.30 for MnO).

Based on this understanding, we can estimate that ionic compounds generally require a greater fraction of exact exchange, and their optimal mixing parameters are greater or equal to the standard mixing parameter ($\alpha = 0.25$), which is in agreement with previous computational results: Han *et al.* reported that MgO, a prototypical ionic oxide, is best described using a high mixing parameter of 0.38,[53] and Siegel *et al.*, showed that the band structure of $Li_2O_2$, a strongly ionic compound, is only well described with a high mixing parameter of 0.48.[55] In contrast, strongly covalent compounds, such as TM sulfides, which possess lower $\Delta$ (usually below 4.0 eV),[83] call for mixing parameters $\alpha < 0.25$. The various TM-O bond lengths in different TM oxide materials provide an intuitive estimate of their covalency, i.e., an increasing TM-O bond length can be interpreted as reduction of the covalent bond character (requiring a larger fraction of exact exchange). Therefore, the optimal mixing parameter of rocksalt-type cation-disordered Li-TM oxides,[22] of which the TM-O bond length is longer than that of the ordered (layered) Li-TM oxides, may be higher than corresponding ordered Li-TM oxides.

As the covalency generally increases with the oxidation state,[70] the ideal fraction of exact exchange for MO ($M^{2+}$) should be greater than the one for $LiMO_2$ ($M^{3+}$), which is in agreement with our predictions. However, we find that the optimal mixing parameter for $Li_{1-x}CoO_2$ increases from 0.17 for the fully lithiated material ($LiCoO_2$) to 0.24 upon complete delithiation ($CoO_2$), even though Co is more oxidized in the latter state. The origin of this trend could be the rehybridization of Co and oxygen states that occurs simultaneously with a local structural distortion of the Co-O bond during delithiation, and which results in a decrease of the Co-O bond covalency.[27]

In the previous section we demonstrated that the average voltages of LiCoO$_2$ and LiNiO$_2$ calculated with the HSE functional using optimized mixing parameters are in excellent agreement with experimental values (**Figures 5, 6 and 7**). We therefore conclude that the HSE functional with proper mixing parameter predicts accurate ground state energies and electronic structures. The results also show that HSE with optimal mixing parameter predicts more accurate redox potentials for Li-TM-oxides than GGA, GGA+$U$ and standard HSE06. Note that the average voltage increases linearly with the fraction of exact HF exchange (**Figures 5 and 7**), as the electrons are more localized on the TM and oxygen atoms with higher mixing parameters. Hence, the covalency of the M-O bond decreases with increasing mixing parameter (see above), which decreases the energy og the metal states and in turn increases the redox potential increases.

Apart from controlling the band gap, admixing HF exchange to GGA has a delicate impact on the relative position of the energy levels of the Co 3$d$ and O 2$p$ states near the Fermi level, which in turn determines the strength of the hybridization between those states. As a result, the ratio of the O 2$p$ states to the Co 3$d$ states near the Fermi level varies strongly with the mixing parameter (**Figures 8 and 9).** As the ratio increases with the mixing parameter, the hybridization between O 2$p$ orbitals and Co $d$ orbitals becomes stronger. The O 2$p$/Co 3$d$ ratio predicted by optimized HSE ($\alpha = 0.17$) is much greater than that of GGA ($\alpha = 0$), which implies that the hybridization between the Co 3$d$ orbitals and the O 2$p$ orbitals becomes stronger than what is predicted by GGA. Indeed, Galakhov *et al.* showed, using Co-L$\alpha$ and O-K$\alpha$ X-ray emission spectroscopy, that the Co 3$d$ and O 2$p$ states are strongly hybridized in LiCoO$_2$.[81] Note that GGA+$U$ predicts a far greater O 2$p$/Co 3$d$ ratio than the optimized HSE functional ($\alpha = 0.17$), thus hybridization between the Co $3d$ orbitals and the O $2p$ orbitals is overestimated. This may explain why GGA+$U$ predicts the wrong average voltage and voltage profile of $LiCoO_2$ (**Figure 6**) even though the GGA+$U$ band gap and O K-edge EELS are similar to the experimental results (**Figure 3**). As this hybridization becomes stronger, the participation of oxygen in the charge compensation upon Li extraction from $LiCoO_2$ also increases. Therefore, the hybrid functional mixing parameter has to be optimized to investigate CT in Li-TM-oxides during lithium deintercalation. However, the computational cost to calculate band gaps with many different mixing parameters is significant. We therefore propose an alternative method to determine the optimal mixing parameter for each system: the band gap of $Li_xCoO_2$ and $Li_xNiO_2$ increases linearly with the amount of exact HF exchange energy within $0 \leq \alpha \leq$ 0.3, as shown in **Figure 4**. This tendency was also observed in our results for MnO and NiO and has previously been reported for $CeO_x$ systems.[54] Thus, the optimal mixing parameter can be obtained by comparing a reference band gap with the linear interpolated band gap between GGA ($\alpha = 0$) and HSE06 ($\alpha = 0.25$). The band gap predicted by $G_0W_0$@GGA+$U$ calculations is, for all considered materials, close to the experimental one (**Figures 1 and 3**) and can thus be used as a reference to determine suitable mixing parameters where experimental data is not available. Nevertheless, care is needed when following this approach, as it is well known that $G_0W_0$ band gaps depend on the starting wave function (GGA+$U$) and thus indirectly depend on the selected Hubbard $U$.[48] When data from XPS-BIS spectra is used as reference, it is crucial that the computed band gap is obtained in the same fashion (with the same resolution) as the experimental one, as discussed for the example of MnO in the previous section.

## CONCLUSION

We propose a methodology for the accurate prediction of electronic structure properties of TM oxides and Li-TM-oxides based on hybrid-functional calculations with optimized mixing parameters. We demonstrated how structure-specific mixing parameters of the HSE functional can be obtained by calibration against experimental and/or $G_0W_0$ band gaps. While the optimized mixing parameters for most TM oxides were found to be close to the standard HSE06 value of 0.25, we observed lower values for Li-TM-oxides. Comparison of computational EELS spectra to experimental references from the literature confirmed that the electronic structures of TM oxides were well reproduced with HSE functional and optimal mixing parameters. The voltage profile for $LiCoO_2$ calculated with HSE and optimal mixing parameter showed clearly improved redox potential as compared to calculations based on GGA($+U$) and standard HSE06. The systematic approach to electronic structure prediction described in this article provides a reliable foundation for the study of subtle electronic structure effects that critically depend on state alignment, such as oxygen redox activity in Li-excess cathode materials or charge-transfer phenomena in semiconductors.

## ACKNOWLEDGEMENT

The authors thank J. Neaton for valuable discussions. This work was supported by the Assistant Secretary for Energy Efficiency and Renewable Energy, Office of Vehicle Technologies of the U.S. Department of Energy under Contract No. DE-AC02–05CH11231, under the Batteries for Advanced Transportation Technologies (BATT) Program subcontract #7056411. Computational resources from the National Energy Research Scientific Computing Center (NERSC) and from the Extreme Science and Engineering Discovery Environment (XSEDE) are gratefully acknowledged. D. H. S. acknowledges the support by Basic Science Research Program through the National Research Foundation of Korea (NRF) funded by the Ministry of Education (2014R1A6A3A03056034).

**Table 1**. Optimal mixing parameters for TM oxides (MO, M = Mn, Co, and Ni), lithium TM oxides ($LiCoO_2$ and $LiNiO_2$) and delithiated lithium TM oxides ($CoO_2$ and $NiO_2$). The mixing parameters of MnO, CoO, NiO and $LiCoO_2$ were optimized against the experimentally observed density of states, whereas those of $CoO_2$, $LiNiO_2$, $NiO_2$ were optimized against $G_0W_0$@GGA+U band gaps.

|  | Oxidation state | Electronic configuration | Optimal mixing parameter |
|---|---|---|---|
| MnO | 2+ | $t_{2g}^{3} e_{g}^{2}$ | 0.30 |
| CoO | 2+ | $t_{2g}^{6} e_{g}^{1}$ | 0.20 |
| NiO | 2+ | $t_{2g}^{6} e_{g}^{2}$ | 0.25 |
| $LiCoO_2$ | 3+ | $t_{2g}^{6} e_{g}^{0}$ | 0.17 |
| $CoO_2$ | 4+ | $t_{2g}^{5} e_{g}^{0}$ | 0.24 |
| $LiNiO_2$ | 3+ | $t_{2g}^{6} e_{g}^{1}$ | 0.18 |
| $NiO_2$ | 4+ | $t_{2g}^{6} e_{g}^{0}$ | 0.25 |

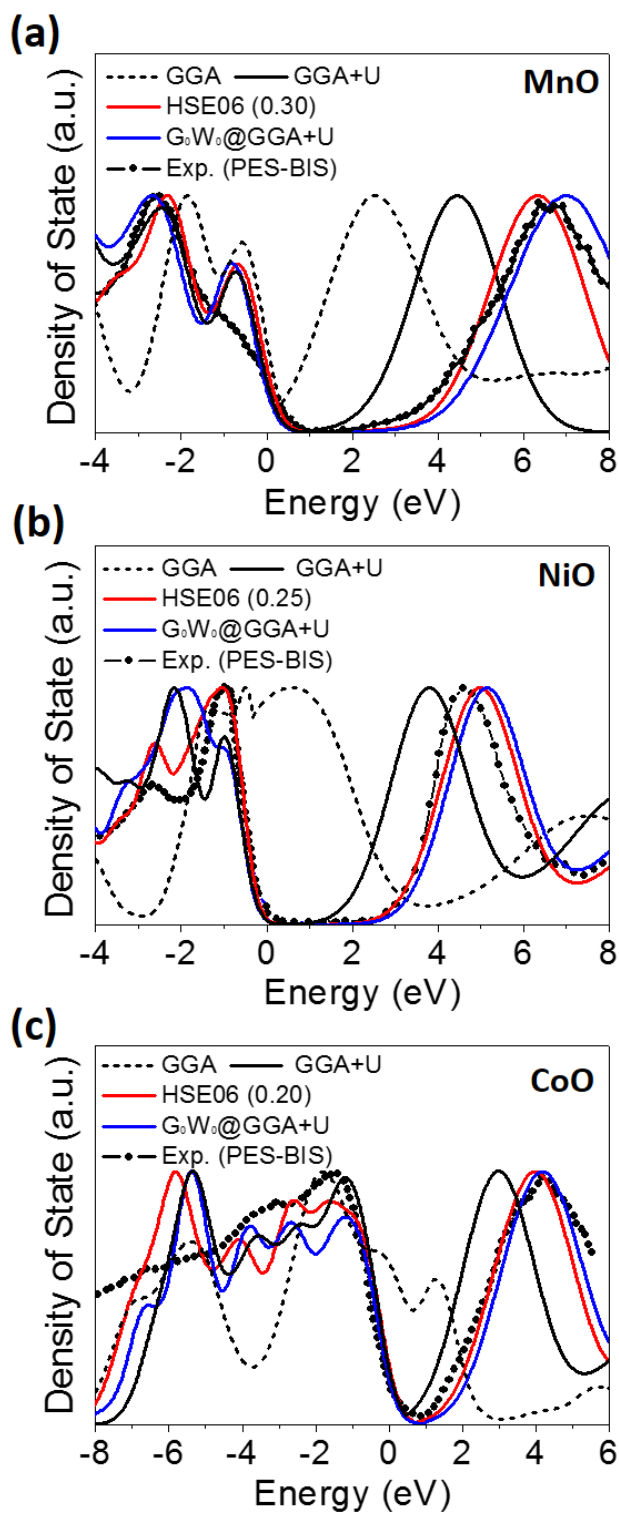

**Figure 1**. Density of states (DOS) of (a) MnO, (b) NiO and (c) CoO as predicted by GGA, GGA+U, HSE with optimal mixing parameter, and $G_0W_0$@GGA+U in comparison to the experimental reference (PES-BIS).

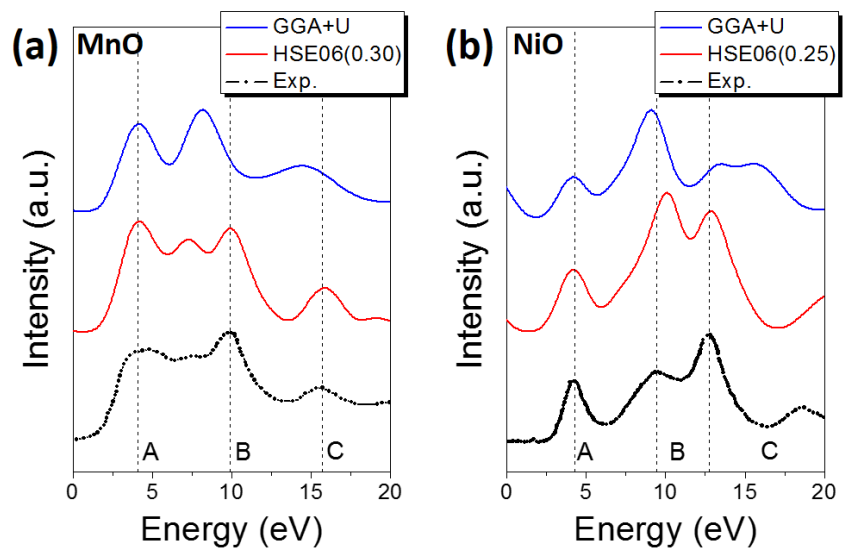

**Figure 2.** Computed and experimental O K-edge EELS spectra of (a) MnO and (b) NiO (GGA+U, HSE06 with optimal mixing parameter, and experimental reference)

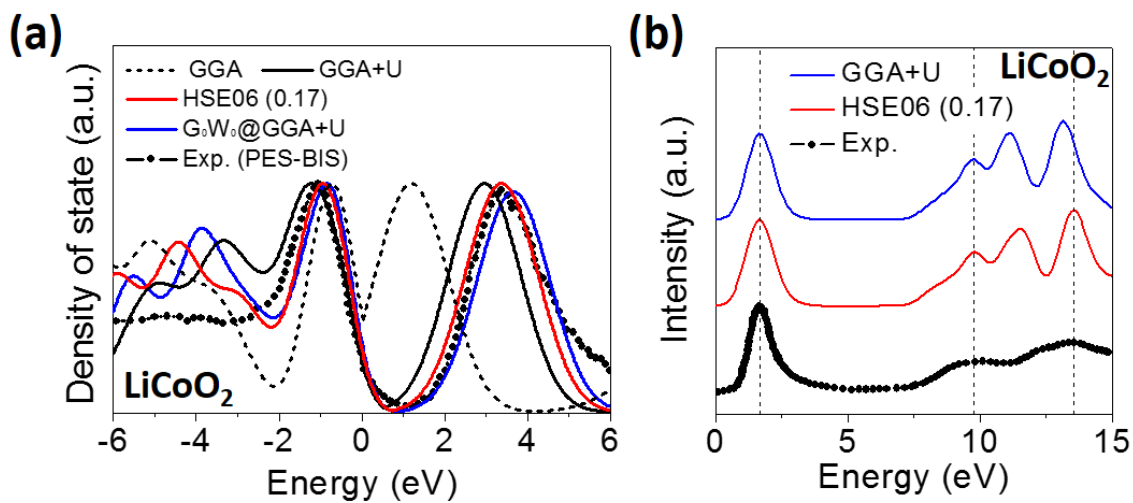

**Figure 3.** (a) Density of states (DOS) of LiCoO$_2$ and (b) O K-edge EELS spectra of LiCoO$_2$ as predicted by various electronic structure methods in comparison to the experimental references.

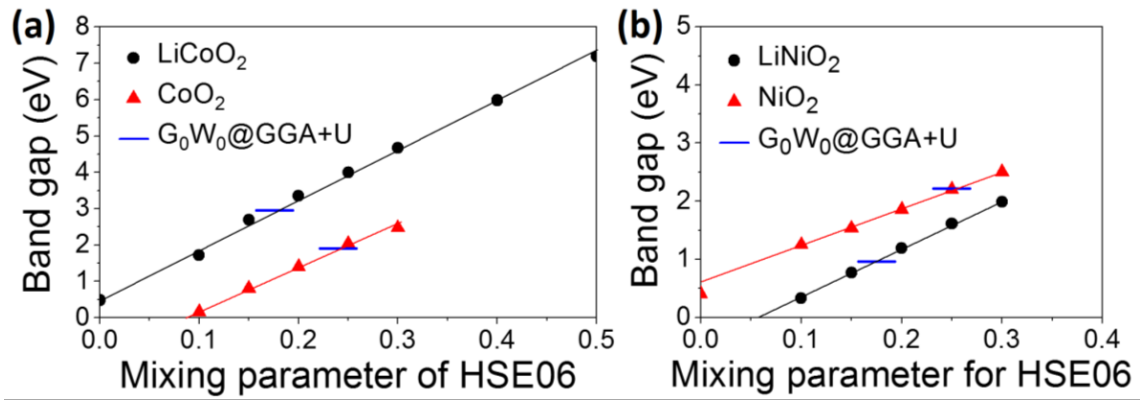

**Figure 4**. Band gaps of (a) LiCoO$_2$ and CoO$_2$ and (b) LiNiO$_2$ and NiO$_2$ as predicted by HSE with increasing mixing parameter. The short (blue) horizontal lines indicate band gaps calculated with G$_0$W$_0$@GGA+U. The solid lines indicate the linear trend of band gaps with increasing fraction of exact HF exchange.

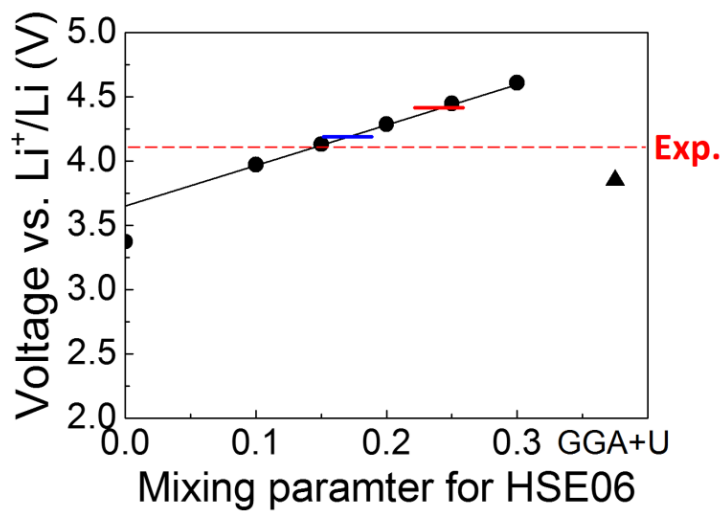

**Figure 5**. Average voltage of Li$_{1-x}$CoO$_2$ ($0 \leq x \leq 1$) as a function of the HSE mixing parameter. The (blue and red) short horizontal lines indicate the voltages calculated with the optimal mixing parameters of LiCoO$_2$ (0.17) and CoO$_2$ (0.24). The (red) dashed line indicates the experimental average voltage of Li$_{1-x}$CoO$_2$ ($0 \leq x \leq 1$), and the black line indicates the linear trend with increasing fraction of exact HF exchange.

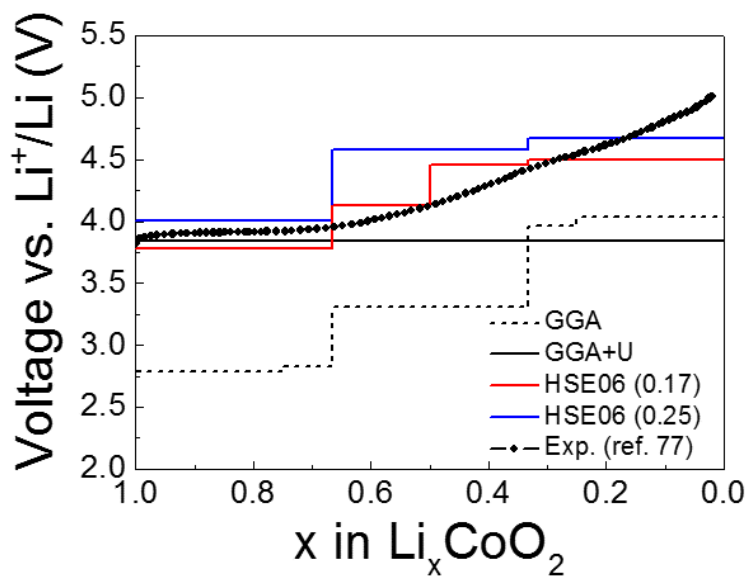

**Figure 6**. Computed voltage profiles of $Li_{1-x}CoO_2$ ($0 \leq x \leq 1$), as predicted by GGA, GGA+U, and HSE with different mixing parameters in comparison to the experimental reference.

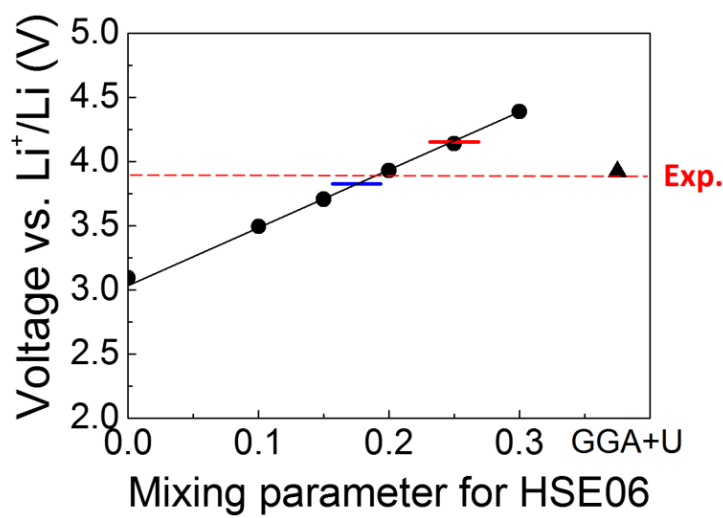

**Figure 7**. Average voltage of $Li_{1-x}NiO_2$ ($0 \leq x \leq 1$) as a function of alpha value. The dashed (red) line indicates the experimental average voltage of $Li_{1-x}NiO_2$ ($0 \leq x \leq 1$) and the solid (black) line indicates the linear trend with increasing fraction of exact exchange. The short (blue and red) horizontal lines indicate the voltages calculated with the optimal mixing parameters of $LiNiO_2$ (0.18) and $NiO_2$ (0.25).

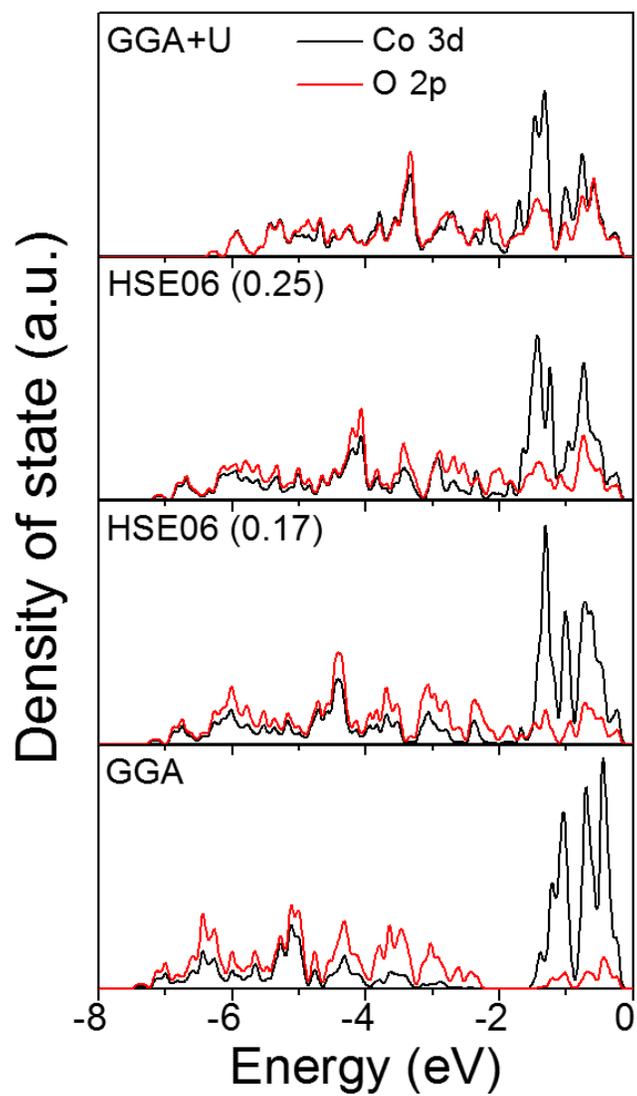

**Figure 8**, Projected density of states (pDOS) of the Co 3*d* orbitals [black] and O 2*p* orbitals [red] in $LiCoO_2$ predicted by GGA, HSE06 ($\alpha = 0.17$), HSE06 ($\alpha = 0.25$), and GGA+U. The Fermi energy is located at 0 eV.

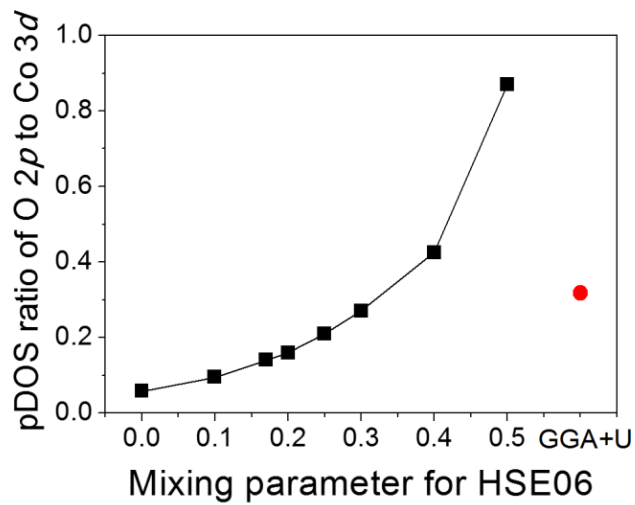

**Figure 9**. Ratio of the O 2*p* pDOS to the Co 3*d* pDOS (black square) in the energy range from 0 eV to -2 eV as a function of the mixing parameter. The ratio predicted by GGA+U (red circle) is also shown.

## References

[1]     G. Ceder, Y. M. Chiang, D. R. Sadoway, M. K. Aydinol, Y. I. Jang, and B. Huang, Nature **392**, 694 (1998).

[2]     M. K. Aydinol, A. F. Kohan, G. Ceder, K. Cho, and J. Joannopoulos, Phys. Rev. B **56**, 1354 (1997).

[3]     M. K. Aydinol, A. F. Kohan, and G. Ceder, J. Power Sources **68**, 664 (1997).

[4]     M. Sathiya, G. Rousse, K. Ramesha, C. P. Laisa, H. Vezin, M. T. Sougrati, M. L. Doublet, D. Foix, D. Gonbeau, W. Walker, A. S. Prakash, M. Ben Hassine, L. Dupont, and J. M. Tarascon, Nat. Mater. **12**, 827 (2013).

[5]     C. F. Petersburg, Z. Li, N. A. Chernova, M. S. Whittingham, and F. M. Alamgir, J. Mater. Chem. **22**, 19993 (2012).

[6]     J. Suntivich, H. A. Gasteiger, N. Yabuuchi, H. Nakanishi, J. B. Goodenough, and Y. Shao-Horn, Nat Chem **3**, 546 (2011).

[7]     J. Suntivich, K. J. May, H. A. Gasteiger, J. B. Goodenough, and Y. Shao-Horn, Science **334**, 1383 (2011).

[8]     T. Arima, Y. Tokura, and J. B. Torrance, Phys. Rev. B **48**, 17006 (1993).

[9]     B. B. Van Aken, T. T. M. Palstra, A. Filippetti, and N. A. Spaldin, Nat. Mater. **3**, 164 (2004).

[10]    S.-W. Cheong and M. Mostovoy, Nat. Mater. **6**, 13 (2007).

[11]    Y. Kamihara, T. Watanabe, M. Hirano, and H. Hosono, J. Am. Chem. Soc. **130**, 3296 (2008).

[12]    J. Zaanen, G. A. Sawatzky, and J. W. Allen, Phys. Rev. Lett. **55**, 418 (1985).

[13]    M. Imada, A. Fujimori, and Y. Tokura, Rev. Mod. Phys. **70**, 1039 (1998).

[14]    J. van Elp, R. H. Potze, H. Eskes, R. Berger, and G. A. Sawatzky, Phys. Rev. B **44**, 1530 (1991).

[15]    J. van Elp, J. L. Wieland, H. Eskes, P. Kuiper, G. A. Sawatzky, F. M. F. de Groot, and T. S. Turner, Phys. Rev. B **44**, 6090 (1991).

[16]    A. Ito, Y. Sato, T. Sanada, M. Hatano, H. Horie, and Y. Ohsawa, J. Power Sources **196**, 6828 (2011).

[17]    H. Koga, L. Croguennec, M. Ménétrier, P. Mannessiez, F. Weill, and C. Delmas, J. Power Sources **236**, 250 (2013).

[18]    M. Sathiya, K. Ramesha, G. Rousse, D. Foix, D. Gonbeau, A. S. Prakash, M. L. Doublet, K. Hemalatha, and J. M. Tarascon, Chem. Mater. **25**, 1121 (2013).

[19]    M. Sathiya, A. M. Abakumov, D. Foix, G. Rousse, K. Ramesha, M. Saubanère, M. L. Doublet, H. Vezin, C. P. Laisa, A. S. Prakash, D. Gonbeau, G. VanTendeloo, and J. M. Tarascon, Nat. Mater. **14**, 230 (2014).

[20]    S.-I. Okuoka, Y. Ogasawara, Y. Suga, M. Hibino, T. Kudo, H. Ono, K. Yonehara, Y. Sumida, Y. Yamada, A. Yamada, M. Oshima, E. Tochigi, N. Shibata, Y. Ikuhara, and N. Mizuno, Sci. Rep. **4**, 5684 (2014).

[21]    N. Twu, X. Li, A. Urban, M. Balasubramanian, J. Lee, L. Liu, and G. Ceder, Nano Lett. **15**, 596 (2015).

[22]    N. Yabuuchi, M. Takeuchi, M. Nakayama, H. Shiiba, M. Ogawa, K. Nakayama, T. Ohta, D. Endo, T. Ozaki, T. Inamasu, K. Sato, and S. Komaba, Proceedings of the National Academy of Sciences, online published (2015).

[23]    R. Wang, X. Li, L. Liu, J. Lee, D.-H. Seo, S.-H. Bo, A. Urban, and G. Ceder, in preparation.

[24]    M. M. Thackeray, P. J. Johnson, L. A. de Picciotto, P. G. Bruce, and J. B. Goodenough, Materials Research Bulletin **19**, 179 (1984).

[25]    A. K. Padhi, K. S. Nanjundaswamy, and J. B. Goodenough, J. Electrochem. Soc. **144**, 1188 (1997).

[26]    A. Van der Ven, M. K. Aydinol, G. Ceder, G. Kresse, and J. Hafner, Phys. Rev. B **58**, 2975 (1998).

[27]    W.-S. Yoon, K.-B. Kim, M.-G. Kim, M.-K. Lee, H.-J. Shin, J.-M. Lee, and J.-S. Lee, The Journal of Physical Chemistry B **106**, 2526 (2002).

[28]    J. Graetz, A. Hightower, C. C. Ahn, R. Yazami, P. Rez, and B. Fultz, The Journal of Physical Chemistry B **106**, 1286 (2002).

[29]    W. Kohn and L. J. Sham, Phys. Rev. **140**, A1133 (1965).

[30]    V. I. Anisimov, J. Zaanen, and O. K. Andersen, Phys. Rev. B **44**, 943 (1991).

[31]    A. I. Liechtenstein, V. I. Anisimov, and J. Zaanen, Phys. Rev. B **52**, R5467 (1995).

[32]    V. I. Anisimov, F. Aryasetiawan, and A. I. Lichtenstein, J. Phys.: Condens. Matter **9**, 767 (1997).

[33]    M. Nolan and G. W. Watson, J. Chem. Phys. **125**, 144701 (2006).

[34]    M. Nolan and G. W. Watson, Surf. Sci. **586**, 25 (2005).

[35]    T. Bredow and A. R. Gerson, Phys. Rev. B **61**, 5194 (2000).

[36]    M. Cococcioni and S. de Gironcoli, Phys. Rev. B **71**, 035105 (2005).

[37]    C. Franchini, V. Bayer, R. Podloucky, J. Paier, and G. Kresse, Phys. Rev. B **72**, 045132 (2005).

[38]    J.-R. Moret, Master thesis, École Polytechnique Fédérale de Lausanne, 2009.

[39]    C. Rödl, F. Fuchs, J. Furthmüller, and F. Bechstedt, Phys. Rev. B **79**, 235114 (2009).

[40]    R. Gillen and J. Robertson, J. Phys.: Condens. Matter **25**, 165502 (2013).

[41]    F. Zhou, M. Cococcioni, C. A. Marianetti, D. Morgan, and G. Ceder, Phys. Rev. B **70**, 235121 (2004).

[42]    A. D. Becke, J. Chem. Phys. **98**, 5648 (1993).

[43]    V. L. Chevrier, S. P. Ong, R. Armiento, M. K. Y. Chan, and G. Ceder, Phys. Rev. B **82**, 075122 (2010).

[44]    S. P. Ong, Y. Mo, and G. Ceder, Phys. Rev. B **85**, 081105 (2012).

[45]    J. Heyd, G. E. Scuseria, and M. Ernzerhof, J. Chem. Phys. **118**, 8207 (2003).

[46]    J. Heyd and G. E. Scuseria, J. Chem. Phys. **121**, 1187 (2004).

[47]    J. Heyd, J. E. Peralta, G. E. Scuseria, and R. L. Martin, J. Chem. Phys. **123**, 174101 (2005).

[48]    H. Jiang, R. I. Gomez-Abal, P. Rinke, and M. Scheffler, Phys. Rev. B **82**, 045108 (2010).

[49]    P. Stephens and F. Devlin, The Journal of Physical Chemistry **98**, 11623 (1994).

[50]    A. Görling and M. Levy, J. Chem. Phys. **106**, 2675 (1997).

[51]    C. Adamo and V. Barone, J. Chem. Phys. **110**, 6158 (1999).

[52]    M. S. Hybertsen and S. G. Louie, Phys. Rev. B **34**, 5390 (1986).

[53]    S. Park, B. Lee, S. H. Jeon, and S. Han, Current Applied Physics **11**, S337 (2011).

[54]    J. s. Graciani, A. M. Márquez, J. J. Plata, Y. Ortega, N. C. Hernández, A. Meyer, C. M. Zicovich-Wilson, and J. F. Sanz, J. Chem. Theory Comput. **7**, 56 (2010).

[55]    M. D. Radin and D. J. Siegel, Energy Environ. Sci. **6**, 2370 (2013).

[56]    J. P. Perdew, K. Burke, and M. Ernzerhof, Phys. Rev. Lett. **77**, 3865 (1996).

[57]    G. Kresse and J. Furthmuller, Comput. Mater. Sci. **6**, 15 (1996).

[58]    H. J. Orman and P. J. Wiseman, Acta Crystallograph. Sect. C **40**, 12 (1984).

[59]    H. Chen, C. L. Freeman, and J. H. Harding, Phys. Rev. B **84**, 085108 (2011).

[60]    S. L. Dudarev, G. A. Botton, S. Y. Savrasov, C. J. Humphreys, and A. P. Sutton, Phys. Rev. B **57**, 1505 (1998).

[61]    A. Jain, G. Hautier, S. P. Ong, C. J. Moore, C. C. Fischer, K. A. Persson, and G. Ceder, Phys. Rev. B **84**, 045115 (2011).

[62]    J. Heyd, G. E. Scuseria, and M. Ernzerhof, J. Chem. Phys. **124**, 219906 (2006).

[63]    J. Paier, M. Marsman, K. Hummer, G. Kresse, I. C. Gerber, and J. G. Ángyán, J. Chem. Phys. **125**, 249901 (2006).

[64]    L. Hedin, Phys. Rev. **139**, A796 (1965).

[65]    P. Liao and E. A. Carter, Phys. Chem. Chem. Phys. **13**, 15189 (2011).

[66]    N. Alidoust, M. C. Toroker, J. A. Keith, and E. A. Carter, ChemSusChem **7**, 195 (2014).

[67]    R. Buczko, G. Duscher, S. J. Pennycook, and S. T. Pantelides, Phys. Rev. Lett. **85**, 2168 (2000).

[68]    C. Elsässer and S. Köstlmeier, Ultramicroscopy **86**, 325 (2001).

[69]    G. A. Sawatzky and J. W. Allen, Phys. Rev. Lett. **53**, 2339 (1984).

[70]    R. Zimmermann, P. Steiner, R. Claessen, F. Reinert, S. Hüfner, P. Blaha, and P. Dufek, J. Phys.: Condens. Matter **11**, 1657 (1999).

[71]    H. Kurata and C. Colliex, Phys. Rev. B **48**, 2102 (1993).

[72]    L. V. Dobysheva, P. L. Potapov, and D. Schryvers, Phys. Rev. B **69**, 184404 (2004).

[73]    V. R. Galakhov, M. Neumann, and D. G. Kellerman, Applied Physics A **94**, 497 (2009).

[74]    G. G. Amatucci, J. M. Tarascon, and L. C. Klein, J. Electrochem. Soc. **143**, 1114 (1996).

[75]    A. Van der Ven, M. K. Aydinol, G. Ceder, G. Kresse, and J. Hafner, Phys. Rev. B **58**, 2975 (1998).

[76]    G. Ceder and A. Van der Ven, Electrochim. Acta **45**, 131 (1999).

[77]    T. Okumura, Y. Yamaguchi, M. Shikano, and H. Kobayashi, J. Mater. Chem. **22**, 17340 (2012).

[78]    M. Menetrier, I. Saadoune, S. Levasseur, and C. Delmas, J. Mater. Chem. **9**, 1135 (1999).

[79]    C. A. Marianetti, G. Kotliar, and G. Ceder, Nat. Mater. **3**, 627 (2004).

[80]    C. Delmas, M. Ménétrier, L. Croguennec, S. Levasseur, J. P. Pérès, C. Pouillerie, G. Prado, L. Fournès, and F. Weill, International Journal of Inorganic Materials **1**, 11 (1999).

[81]    V. R. Galakhov, E. Z. Kurmaev, S. Uhlenbrock, M. Neumann, D. G. Kellerman, and V. S. Gorshkov, Solid State Commun. **99**, 221 (1996).

[82]    J. van Elp, H. Eskes, P. Kuiper, and G. A. Sawatzky, Phys. Rev. B **45**, 1612 (1992).

[83]    A. E. Bocquet, T. Mizokawa, T. Saitoh, H. Namatame, and A. Fujimori, Phys. Rev. B **46**, 3771 (1992).